\pdfoutput=1

\documentclass[11pt]{article}
\usepackage[acceptedWithA]{tacl2021v1}

\usepackage[T1]{fontenc}

\usepackage{times}
\usepackage{url}
\usepackage{latexsym}
\usepackage{enumitem}
\usepackage{color, colortbl}
\usepackage{threeparttable,booktabs}
\usepackage{amssymb}
\usepackage{graphicx,subcaption}

\usepackage[hang,flushmargin]{footmisc}

\usepackage{microtype}

\usepackage{booktabs}
\usepackage{multirow}
\usepackage{adjustbox}
\usepackage{arydshln}
\usepackage{footnote}
\usepackage{tabularx}
\usepackage{array}
\usepackage{graphicx,subcaption}
\usepackage{enumitem}
\usepackage{amsmath}
\usepackage{amsfonts}
\usepackage{dsfont}
\usepackage{pifont}
\usepackage{xspace}

\definecolor{bluencs}{rgb}{0.0, 0.53, 0.74}
\newcommand{\cmark}{\ding{51}}%
\newcommand{\xmark}{\ding{55}}%

\newcommand{\aggretriever}{Aggretriever\xspace}
\newcommand{\clsbert}{\text{BERT}_{\text{CLS}}}
\newcommand{\clsdistilbert}{\text{DistilBERT}_{\text{CLS}}}

\newcommand{\clscondenser}{\text{Condenser}_{\text{CLS}}}
\newcommand{\clscocondenser}{\text{coCondenser}_{\text{CLS}}}
\newcommand{\aggbert}{\text{BERT}_{\text{AGG}}}
\newcommand{\aggdistilbert}{\text{DistilBERT}_{\text{AGG}}}

\newcommand{\aggcondenser}{\text{Condenser}_{\text{AGG}}}
\newcommand{\aggcocondenser}{\text{coCondenser}_{\text{AGG}}}

\definecolor{Gray}{gray}{0.9}

\title{\aggretriever: A Simple Approach to Aggregate Textual Representations for Robust Dense Passage Retrieval}

\author{Sheng-Chieh Lin, Minghan Li \and Jimmy Lin\\[1ex]
David R. Cheriton School of Computer Science\\University of Waterloo\\[1ex]
        \texttt{\{s269lin,m692li,jimmylin\}@uwaterloo.ca}}

\begin{document}
\maketitle
\begin{abstract}
Pre-trained language models have been successful in many knowledge-intensive NLP tasks.
However, recent work has shown that models such as BERT are not ``structurally ready'' to aggregate textual information into a \texttt{[CLS]} vector for dense passage retrieval (DPR).
This ``lack of readiness'' results from the gap between language model pre-training and DPR fine-tuning. 
Previous solutions call for computationally expensive techniques such as hard negative mining, cross-encoder distillation, and further pre-training to learn a robust DPR model.
In this work, we instead propose to fully exploit knowledge in a pre-trained language model for DPR by aggregating the contextualized token embeddings into a dense vector, which we call $\mathbf{agg}^{\star}$. 
By concatenating vectors from the \texttt{[CLS]} token and $\mathbf{agg}^{\star}$, our \textit{\aggretriever} model substantially improves the effectiveness of dense retrieval models on both in-domain and zero-shot evaluations without introducing substantial training overhead.
Code is available at \url{https://github.com/castorini/dhr}

\end{abstract}

\maketitle

\section{Introduction}
\label{sec:intro}

A bi-encoder architecture~\cite{sentence-bert, dpr} based on pre-trained language models~\cite{devlin2018bert, liu2019roberta, raffel2019t5} has been widely used for first-stage retrieval in knowledge-intensive tasks such as open-domain question answering and fact checking.
Compared to bag-of-words models such as BM25, these approaches circumvent lexical mismatches between queries and passages by encoding text into dense vectors. 

\begin{figure}[t]
    \centering
    \resizebox{1\columnwidth}{!}{
        \includegraphics{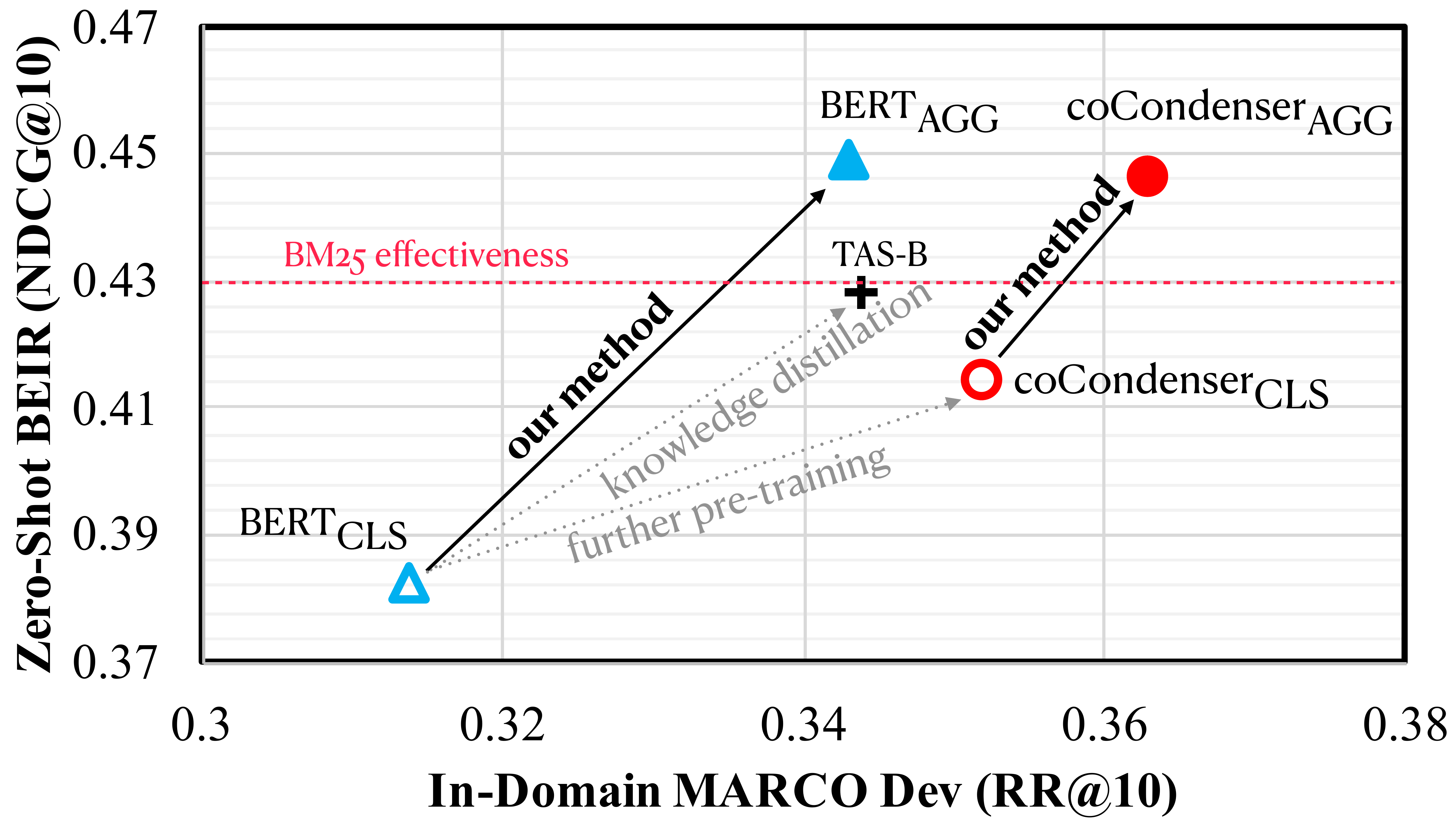}
    }
    \caption{In-domain versus zero-shot effectiveness. All DPR models are trained with BM25 negatives.} 
    \label{fig:teaser}
\end{figure}

Despite their success, recent research calls into question the robustness of these single-vector models~\citep{beir}.  
As shown in Fig.~\ref{fig:teaser}, single-vector dense retrievers (e.g., $\clsbert$ and TAS-B) trained with well-designed knowledge distillation strategies~\citep{tasb} still underperform BM25 on out-of-domain datasets. 
Along the same lines, \citet{sciavolino2021simple} find that simple entity-centric questions are challenging to these dense retrievers. 

\begin{figure*}[t]
    \centering
    \resizebox{1\textwidth}{!}{
        \includegraphics{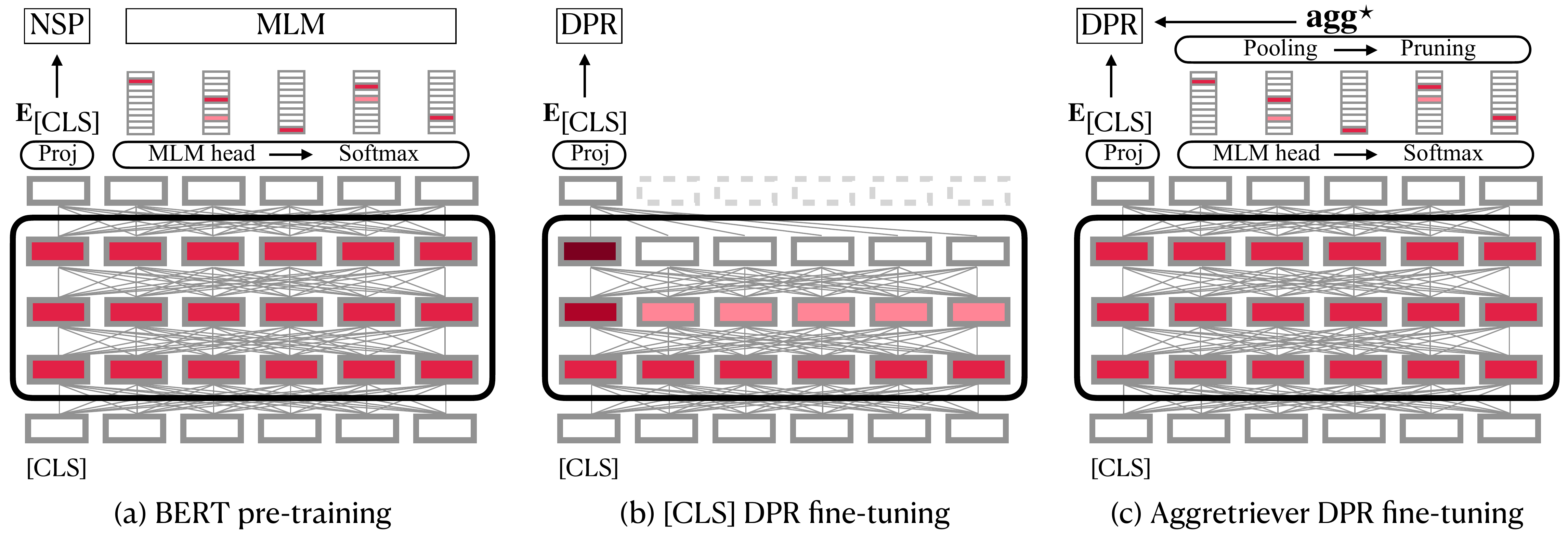}
    }
    \caption{(a) BERT: next sentence prediction (NSP) and mask language modeling (MLM) (b) DPR: using the \texttt{[CLS]} embedding for retrieval (c) \aggretriever: aggregating knowledge from both NSP and MLM.} 
    \label{fig:dpr_comparison}
\end{figure*}

Recently, \citet{condenser} observe that pre-trained language models such as BERT are not ``structurally ready'' for fine-tuning on downstream retrieval tasks. 
This is because the \texttt{[CLS]} token, pre-trained on the task of next sentence prediction (NSP), does not have the proper attention structure to aggregate fine-grained textual information. 
To address this issue, the authors propose to further pre-train the \texttt{[CLS]} vector before fine-tuning and show that the gap between pre-training and fine-tuning tasks can be mitigated (see $\clscocondenser$ illustrated in Fig.~\ref{fig:teaser}). 
However, further pre-training introduces additional computational costs, which motivates us to ask the following question: Can we directly bridge the gap between pre-training and fine-tuning without any further pre-training? 

Before diving into our proposed solution, we briefly overview the language modeling pre-training and DPR fine-tuning tasks using BERT.
Fig.~\ref{fig:dpr_comparison}(a) illustrates the BERT pre-training tasks, next sentence prediction (NSP) and mask language modeling (MLM), while Fig.~\ref{fig:dpr_comparison}(b) shows the task of fine-tuning a dense retriever.
We observe that solely relying on the \texttt{[CLS]} vector as the dense representation does not exploit the full capacity of the pre-trained model, as the \texttt{[CLS]} vector participates directly only in NSP during pre-training, and therefore lacks information captured in the contextualized token embeddings. 
A simple solution is to aggregate the token embeddings by pooling (max or mean) into a single vector.
However, information is lost in this process and empirical results do not show any consistent effectiveness gains.
Hence, we see the need for better aggregation schemes.

In this paper, we propose a novel approach to generate textual representations for retrieval that fully exploit contextualized token embeddings from BERT, shown in Fig.~\ref{fig:dpr_comparison}(c).
Specifically, we reuse the pre-trained MLM head to map each contextualized token embedding into a high-dimensional wordpiece lexical space. 
Following a simple max-pooling and pruning strategy, we obtain a compact {\it lexical} vector that we call $\mathbf{agg}^{\star}$.
By concatenating $\mathbf{agg}^{\star}$ and the \texttt{[CLS]} vector, our novel \textit{\aggretriever} dense retrieval model captures representations pre-trained from both NSP and MLM, improving retrieval effectiveness by a noticeable margin compared to fine-tuned models that solely rely on the \texttt{[CLS]} vector (see $\aggbert$ vs $\clsbert$ in Fig.~\ref{fig:teaser}). 

Importantly, fine-tuning \aggretriever does not require any sophisticated and computationally expensive techniques, making it a simple yet competitive baseline for dense retrieval. 
However, our approach is orthogonal to previously proposed further pre-training strategies, and can still benefit from them to improve retrieval effectiveness even more (see $\aggcocondenser$ in Fig.~\ref{fig:teaser}). 
To the best of our knowledge, this is the first work in the DPR literature that leverages the BERT pre-trained MLM head to encode textual information into a single dense vector.

\section{Background and Motivation}

Given a query $q$, our task is to retrieve a list of passages to maximize some ranking metric such as nDCG or MRR.
Dense retrievers~\cite{sentence-bert, dpr} based on pre-trained language models encode queries and passages as low dimensional vectors with a bi-encoder architecture and use the dot product between the encoded vectors as the similarity score:
\begin{align}
\label{eq:cls_scoring_function}
    \textrm{sim}_{\texttt{[CLS]}}(q, p) \triangleq  \mathbf{e}_{q_{\texttt{[CLS]}}} \cdot \mathbf{e}_{p_{\texttt{[CLS]}}},
\end{align}
\noindent where $\mathbf{e}_{q_{\texttt{[CLS]}}}$ and $\mathbf{e}_{p_{\texttt{[CLS]}}}$ are the $\texttt{[CLS]}$ vectors at the last layer of BERT~\citep{devlin2018bert}. 
Subsequent work leverages expensive fine-tuning strategies (e.g., hard negative mining, knowledge distillation) to guide models to learn more effective and robust single-vector representations~\cite{xiong2020approximate, star, tct, tasb, rocketqa}.

Recent work~\citep{condenser,seed-encoder} shows that the \texttt{[CLS]} vector remains ``dormant'' in most layers of pre-trained models and fails to adequately aggregate information from the input sequence during pre-training.
Thus, researchers argue that the models are not ``structurally ready'' for fine-tuning.
To tackle this issue, unsupervised contrastive learning has been proposed, which creates pseudo relevance labels from the target corpus to ``prepare'' the \texttt{[CLS]} vector for retrieval.
The most representative technique is the Inverse Cloze
Task \citep[ICT;][]{ict}.
However, since the generated relevance data is noisy, further pre-training with ICT often requires a huge amount of computation due to the need for large batch sizes or other sophisticated training techniques~\cite{Chang2020Pre-training, contriever, gtr}.

Another thread of work~\cite{condenser, seed-encoder} manages to guide transformers to aggregate textual information into the \texttt{[CLS]} vector through auto-encoding. 
This method does not require as much computation as unsupervised contrastive learning but is still much more computationally intensive than fine-tuning. 
For example, \citet{condenser} report that the further pre-training process still requires one week on four RTX 2080 Ti GPUs, while fine-tuning consumes less than one day in the same environment.

Recent work on neural sparse retrievers~\cite{sparterm, splade} projects contextualized token embeddings into a high-dimensional wordpiece lexical space through the BERT pre-trained MLM projector and directly performs retrieval in wordpiece lexical space. 
These models demonstrate that MLM pre-trained weights can be used to learn effective lexical representations for retrieval tasks, a finding that has not been fully explored in the DPR literature. 
Inspired by this work, we explore reusing MLM pre-trained weights for DPR fine-tuning and further combine the \texttt{[CLS]} vector to fully exploit textual information in a pre-trained language model.

\section{\aggretriever}
\label{sec:aggretriever}

In this section, we first introduce our method for text aggregation to form $\mathbf{agg}^{\star}$, which consists of two steps:\ pooling and pruning.
Then, we describe how to concatenate the aggregated text representation $\mathbf{agg}^{\star}$ and \texttt{[CLS]} into a 768-dimensional dense vector for fine-tuning and retrieval.

\subsection{Text Aggregation Pooling}
\label{subsec:text_aggregation:pooling}

The goal of text aggregation is to transform contextualized token embeddings into a single-vector token representation. 
Let the input sequence $q$ denote a tokenized query sequence with a length of $l$, $(\texttt{[CLS]}, q_1, q_2, \cdots q_l, \texttt{[SEP]})$, or alternatively, a passage $p$ of length $m$, $(\texttt{[CLS]}, p_1, p_2, \cdots p_m, \texttt{[SEP]})$.
One simple approach is to directly pool (mean or max) contextualized token embeddings from the final layer. 
Such pooling strategies have been studied in previous work~\cite{sentence-bert}, but do not appear to be consistently more effective than just using the \texttt{[CLS]} token; this is also confirmed in our ablation study (Section~\ref{subsec:ablation}). 

We instead propose to reuse the pre-trained MLM head to project each contextualized token embedding $\mathbf{e}_{q_i}$ into a high-dimensional vector in the wordpiece lexical space:
\begin{align}
\label{eq:mlm_project}
    \mathbf{p}_{q_i} &=
    \text{softmax}(\mathbf{e}_{q_i} \cdot \mathbf{W}_{\text{MLM}} + \mathbf{b}_{\text{MLM}}),
\end{align}
where $\mathbf{e}_{q_i}\in \mathbb{R}^d$, $\mathbf{W}_{\text{MLM}}\in \mathbb{R}^{d\times|\text{V}_{\text{BERT}}|}$, and $\mathbf{b}_{\text{MLM}}\in \mathbb{R}^{|\text{V}_{\text{BERT}}|}$ are the weights of the pre-trained MLM linear projector, and $\mathbf{p}_{q_i} \in \mathbb{R}^{|\text{V}_{\text{BERT}}|}$ is the $i$-th contextualized token represented by a probability distribution over the 30522 tokens of BERT wordpiece vocabulary, $\text{V}_{\text{BERT}}$.
We then perform weighted max pooling for the sequential representations $(\mathbf{p}_{q_1}, \mathbf{p}_{q_2}, \cdots, \mathbf{p}_{q_l})$ to obtain a single-vector lexical representation:
\begin{align}
\label{eq:token_vector}
    \mathbf{v}_{q}[v] &= \max_{i \in (1,2,\cdots,l)} w_i \cdot   {\mathbf{p}}_{q_i}[v],
\end{align}
where $w_i = |\mathbf{e}_{q_i} \cdot \mathbf{W} + \mathbf{b}| \in \mathbb{R}^{1}$ is a positive scalar and $v \in \{1, 2, \cdots, |\text{V}_{\text{BERT}}|\}$; $\mathbf{W}\in \mathbb{R}^{d \times 1}$ and $\mathbf{b}\in \mathbb{R}^1$ are trainable weights. 
Note that the scalar $w_i$ for each token $q_i$ is essential to capture term importance, which $\mathbf{p}_{q_i}$ alone cannot capture since it is normalized by softmax. 
We exclude the \texttt{[CLS]} token embedding at this stage since it is used for next-sentence prediction during pre-training and thus we argue that it does not carry much lexical information.

Our design has three advantages:\ (1) the MLM head with softmax is used for BERT pre-training; thus, the output probabilities can accurately model each contextualized token semantically. 
(2) In contrast to directly pooling contextualized embeddings, important dimensions of the token representations in the high-dimensional space are less likely to overlap, resulting in non-interfering max-pooling~\cite{ultra}. 
(3) Finally, $w_i$ and ${\mathbf{p}}_{q_i}$ disentangle the effects of term importance from the MLM head.
We will study the effectiveness of this design in Section~\ref{subsec:ablation} through ablations. 
Note that compared to previous work on sparse retrieval~\cite{sparterm,splade}, which switches softmax to ReLU to create sparse representations, our design sticks to the original activation function for MLM pre-training and directly outputs 30522-dimensional dense lexical vectors ($\mathbf{v}_{q}$).

\begin{figure}[t]
    \centering
    \resizebox{0.5\textwidth}{!}{        \includegraphics{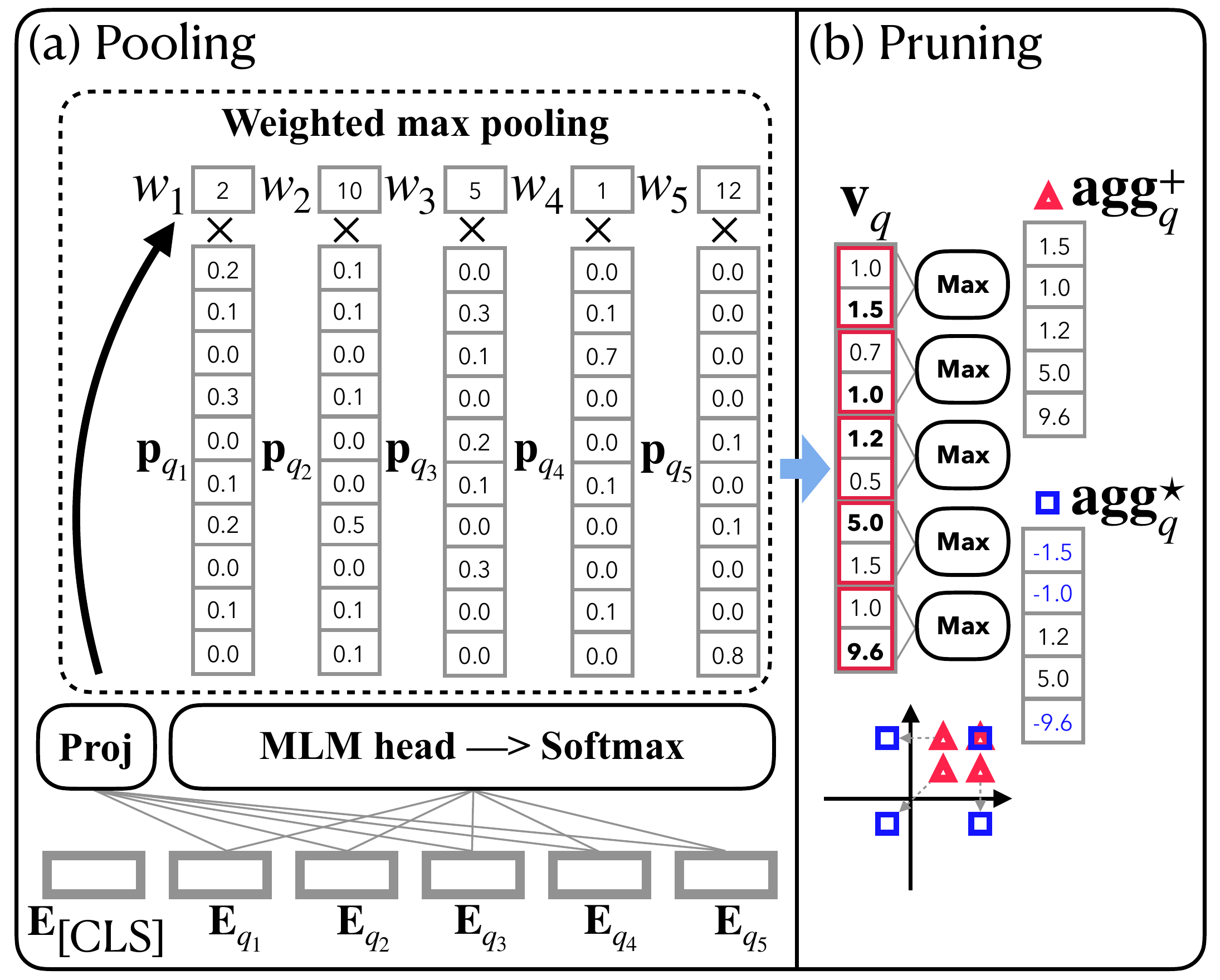}
    }
    \caption{Illustration of text aggregation: (a) pooling of token representations to form $\mathbf{v}_q$; (b) pruning of $\mathbf{v}_q$ to form $\mathbf{agg}^{\star}_{q}$ (or $\mathbf{agg}^{+}_{q}$). While pruning, $\mathbf{agg}^{\star}_{q}[n]$ receives a negative value if the pooled element belongs to $S_n^-$; i.e., the second element in each slice (red box).}
    \label{fig:text_agg_illustration}
\end{figure}

Fig.~\ref{fig:text_agg_illustration}(a) illustrates the generation of $\mathbf{v}_q$ with $|\text{V}_{\text{BERT}}|=10$ for simplicity. 
Ideally, we can directly compute $\mathbf{v}_q \cdot \mathbf{v}_p$ as a lexical matching similarity score for the wordpiece lexical representations.
However, the vectors ($\mathbf{v}_q, \mathbf{v}_p \in \mathbb{R}^{|\text{V}_{\text{BERT}}}|$) are too large for efficient retrieval using dense vector search libraries such as Faiss. 
To address this issue, we introduce our non-parametric pruning method to convert $\mathbf{v}_q$ ($\mathbf{v}_p$) into a low-dimensional vector for dense retrieval. 

\subsection{Text Aggregation Pruning}
\label{subsec:text_aggregation:pruning}

We consider $\mathbf{v}_q \in \mathbb{R}^{|\text{V}_{\text{BERT}}|}$ as a bag-of-words representation with each dimension storing the corresponding term weight.
Thus, dimensions with low term weights indicate that the corresponding terms are not important and can be pruned. 

Based on this intuition, we propose to prune term weights in $\mathbf{v}_q$ by evenly and randomly dividing the dimensions (vocabulary) into $d$ slices, $(S_1, S_2, \cdots, S_d)$, where each slice consists of a set of $\frac{|\text{V}_{\text{BERT}}|}{d}$ index positions.
Then, we condense $\mathbf{v}_q$ into a $d$-dimensional vector by pruning the term weights in each slice $S_n$:
\begin{align}
\label{eq:pruned_token_vector}
    \mathbf{agg}_{q}^+[n] &= \max_{v \in S_n} \ {\mathbf{v}_{q}}[v]; \\
    \mathbf{id}_{q}[n] &= \underset{v \in S_n}{\text{arg}\,\max} \ {\mathbf{v}_{q}}[v]. \nonumber
\end{align}
We call the operation in Eq.~(\ref{eq:pruned_token_vector}) \textit{slice max pooling}, where each value in $\mathbf{agg}^+_q$ represents the weight of the most important term in the slice.\footnote{Slice {\it mean} pooling is less effective in our experiment.}
Slice max pooling is an important operation to prune the term weights while performing dimensionality reduction for dense passage retrieval.
Other effective approaches to pruning lexical representations, e.g., top-$k$ pruning~\cite{top-k_splade} and FLOP regularization~\cite{splade}, do not reduce the vector dimensionality.
Thus, they generate sparse representation models that require inverted indexes for efficient retrieval.

We call $\mathbf{agg}^+_q \in \mathbb{R}^{d}$ the semi-aggregated lexical representation for query $q$ since it only distributes vectors over the positive orthant and does not fully use the $d$-dimensional space.
That is, $\mathbf{v}_q[v] \geq 0 \ \forall \ v \in \{1,2, \cdots |\text{V}_{\text{BERT}}|\}$; thus, $\mathbf{agg}^+_q[n] \geq 0 \ \forall \ n \in \{1,2, \cdots d\}$.
Our goal is to approximate the dot product between $\mathbf{v}_{q}$ and $\mathbf{v}_{p}$ in Eq.~(\ref{eq:token_vector}) by the ones in Eq.~(\ref{eq:pruned_token_vector}):
\begin{align}
\label{eq:approx}
\mathbf{v}_q \cdot \mathbf{v}_p &\approx \sum_{n=1}^{d} (\max_{v \in S_n} \ {\mathbf{v}_{q}}[v]) \cdot (\max_{v \in S_n} \ {\mathbf{v}_{p}}[v]) \nonumber\\
&= \sum_{n=1}^{d} \mathbf{agg}^+_q[n] \cdot \mathbf{agg}^+_p[n] \nonumber\\
&= \mathbf{agg}^+_q \cdot \mathbf{agg}^+_p  .
\end{align}
Note that the approximation error in Eq.~(\ref{eq:approx}) partially comes from \textit{term misalignment}:
\begin{align}
\label{eq:misalignment}
\mathbf{id}_{q}[n] \neq \mathbf{id}_{p}[n],
\end{align}
where the values in $\mathbf{agg}^+_{q}[n]$ and $\mathbf{agg}^+_{p}[n]$ do not represent the same term. 
Alternatively, this can be explained as fuzzy matching between two lexical representations since the two different wordpiece tokens may interact and contribute to the dot product.
Term misalignment increases as $d$ becomes smaller with respect to $|\text{V}_{\text{BERT}}|$; thus, the error increases as well, which we show in Section~\ref{subsec:ablation}.

To mitigate this error, we distribute the semi-aggregated lexical representation to the negative orthants to form what we call the fully aggregated lexical representation, distributed over the entire $d$-dimensional space.
\begin{align}
\label{eq:pruned_token_vector_full}
    \mathbf{agg}^{\star}_q[n] = \begin{cases}
    \ \ \ \mathbf{agg}^+_{q}[n]\ \ \ \text{if } \mathbf{id}_{q}[n] \in S_n^{+}  \\
    -\mathbf{agg}^+_{q}[n]\ \ \ \text{if } \mathbf{id}_{q}[n] \in S_n^{-},
    \end{cases}
\end{align}
where $S_n^+$ and $S_n^-$ are disjoint subsets of $S_n$ (i.e.,
$S_n^+ \cup S_n^- = S_n$ and $S_n^+ \cap S_n^- = \emptyset$). 
That is, we evenly distribute the elements in $S_n$ to $S^+_n$ and $S^-_n$. 

The dot product between two fully aggregated lexical representations then becomes:
\begin{align}
\label{eq:approx_full}
&\textrm{sim}_{\text{agg}}(q, p) \triangleq \mathbf{agg}^{\star}_q \cdot \mathbf{agg}^{\star}_p \\
&= \sum_{n=1}^d
\begin{cases}
-\mathbf{agg}^+_q[n] \cdot \mathbf{agg}^+_p[n] \ \ \text{if case~(a) or (b)}\\
\ \ \ \mathbf{agg}^+_q[n] \cdot \mathbf{agg}^+_p[n] \ \  \text{otherwise,} \nonumber
\end{cases}
\end{align}
where the cases are:
\begin{itemize}
\item[(a)] $\mathbf{id}_{q}[n] \in S_n^{+}; \mathbf{id}_{p}[n] \in S_n^{-}$;
\item[(b)] $\mathbf{id}_{q}[n] \in S_n^{-}; \mathbf{id}_{p}[n] \in S_n^{+}$. 
\end{itemize}
That is, the dot product of $\mathbf{agg}^{\star}$ in Eq.~(\ref{eq:approx_full}) avoids interactions between misaligned terms in the above cases (with 50\% of probability), which $\mathbf{agg}^{+}$ in Eq.~(\ref{eq:approx}) does not consider.  
Note that we do not store the vectors $\mathbf{id}_p$ and $\mathbf{id}_q$ to compute Eq.~(\ref{eq:approx_full}). 
Fig.~\ref{fig:text_agg_illustration}(b) illustrates the difference between $\mathbf{agg}^+_q$ and $\mathbf{agg}^{\star}_q$ with $d=5$, $|S_n|=2$ and $|S^-_n|=|S^+_n|=1$ for simplicity.

\subsection{Fine-Tuning and Retrieval}

Although $\mathbf{agg}^{\star}$ can mitigate the issue of term misalignment, the approximation error cannot be completely eliminated unless $d=|\text{V}_{\text{BERT}}|$. 
To enhance retrieval effectiveness, we concatenate the $\mathbf{agg}^{\star}$ vector with the $\texttt{[CLS]}$ vector since they are pre-trained to capture textual representations in different ways, focusing on the lexical and semantic, respectively.

In our \aggretriever model, the scoring function is the dot product of the concatenated vectors:
\begin{align}
    \textrm{sim}(q, p) \triangleq (\mathbf{e}_{q_{\texttt{[CLS]}}} \oplus \mathbf{agg}_q^{\star}) \cdot (\mathbf{e}_{p_{\texttt{[CLS]}}} \oplus \mathbf{agg}_p^{\star}), \nonumber
\end{align}
where $\oplus$ means vector concatenation.
The vector $\mathbf{e}_{q_{\texttt{[CLS]}}} \oplus \mathbf{agg}_q^{\star}$ captures representations pre-trained from both NSP and MLM. 

During fine-tuning, we minimize the negative log-likelihood of a relevant query--passage pair. 
Specifically, given a query $q$, its relevant passage $p^+$ and a set of negative passages $\{p^-_1, p^-_2, \cdots, p^-_{bs} \}$, we train our model by minimizing the negative log-likelihood (NLL) of the positive $\{q,p^+\}$ pair over all the passages, i.e., $\mathcal{L}$ is
\begin{align}
\label{eq:nnl}
   -\log \frac{{\exp(\textrm{sim}(q, p^+)})}{ \exp({\textrm{sim}(q, p^+)}) + \overset{bs}{\underset{j=1}{\sum}} \exp({\textrm{sim}(q, p^-_j)})}. \nonumber
\end{align}
Following \citet{dpr}, we include the positive and negative passages from the other queries in the same batch as the negatives. 
In addition, we also use the same NLL loss, $\mathcal{L}_{\text{agg}}$ and $\mathcal{L}_{\texttt{[CLS]}}$, to optimize $\textrm{sim}_{\text{agg}}$ and $\textrm{sim}_{\texttt{[CLS]}}$ separately. 
The final loss is as follows:
\begin{align}
    \mathcal{L} + \lambda_1 \cdot \mathcal{L}_{\text{agg}} + \lambda_2 \cdot \mathcal{L}_{\texttt{[CLS]}}. \nonumber
\end{align}
We set $\lambda_1$ and $\lambda_2$ to 0.5 in all our experiments. 
While conducting end-to-end retrieval, we use Flat IP in Faiss~\citep{faiss} to index the passage vectors.
Note that in our main experiments, we project $\mathbf{e}_{q_{\texttt{[CLS]}}}$ and $\mathbf{e}_{p_{\texttt{[CLS]}}}$ to 128 dimensions through a linear layer and set $d=640$ for $\mathbf{agg}^{\star}$ so that the dimensionality is 768.

\begin{table}[t!]
	\caption{Dataset statistics.}
	\label{tb:data_statistics}
	\vspace{-0.3cm}
	\centering
	 \resizebox{1\columnwidth}{!}{
	\setlength\tabcolsep{0.12cm}
    \begin{tabular}{l|r|r|r}
\hline \hline
  \multicolumn{1}{c}{}& \multicolumn{1}{c}{\textbf{MARCO}}& \multicolumn{1}{c}{\textbf{NQ}}& \multicolumn{1}{c}{\textbf{TQA}}\\ 
  \hline
  \arrayrulecolor{lightgray}
\# passages& 8,841,823& \multicolumn{2}{c}{21,015,325}\\ \hline
\arrayrulecolor{black}
\# training queries& 532,761& 58,880& 60,413\\ \hline
\multirow{2}{*}{\# test queries} & \textbf{Dev}\ /\ \textbf{DL19}\ /\ \textbf{20}& \multicolumn{1}{c|}{\textbf{Test}}& \multicolumn{1}{c}{\textbf{Test}} \\\cline{2-4}
& 6,980\ /\ 43\ /\ 53 & 3,610& 11,313\\
\hline \hline
	\end{tabular}}
		\vspace{-0.4cm}
\end{table} 

\section{Experimental Setup}

\begin{table*}[h]
	\caption{In-domain retrieval effectiveness comparisons. All models are fine-tuned with negatives from BM25. Bold denotes the best model for that metric.}
	\label{tb:in_domain_result}
	\centering
	 \resizebox{1\textwidth}{!}{  
	  \setlength\tabcolsep{0.2cm}
    \begin{tabular}{l|ccc|ccc|ccc}
\hline \hline
\multicolumn{1}{c}{}& \multicolumn{2}{c}{\textbf{MARCO Dev}}& \multicolumn{1}{c}{\textbf{DL19}\ /\ \textbf{20}}& \multicolumn{3}{c}{\textbf{\textbf{NQ Test}}}& \multicolumn{3}{c}{\textbf{TQA Test}}\\
 \cmidrule(rl){2-3} \cmidrule(rl){4-4} \cmidrule(rl){5-7} \cmidrule(rl){8-10} 
\multicolumn{1}{l}{Model}& \multicolumn{1}{c}{RR@10}& \multicolumn{1}{c}{R@1K}& \multicolumn{1}{c}{nDCG@10}& \multicolumn{1}{c}{R@5}& \multicolumn{1}{c}{R@20}& \multicolumn{1}{c}{R@100}& \multicolumn{1}{c}{R@5}& \multicolumn{1}{c}{R@20}& \multicolumn{1}{c}{R@100}\\ \hline 
(a) BM25& 0.188& 0.858& 0.506\ /\ 0.475& 0.438& 0.629& 0.783& 0.663& 0.764& 0.832\\ \hline 
 \arrayrulecolor{lightgray}
(1) $\clsdistilbert$& 0.308& 0.940& 0.633\ /\ 0.629& 0.660& 0.785& 0.860& 0.698& 0.790& 0.849\\
(2) $\aggdistilbert$& 0.341& 0.960& \textbf{0.682}\ /\ 0.674& 0.681& 0.805& 0.869& 0.729& 0.808& 0.857\\ \cline{1-10} 
(3) $\clsbert$& 0.314& 0.942& 0.612\ /\ 0.643& 0.677& 0.799& 0.863& 0.710& 0.796& 0.852\\
  \arrayrulecolor{black}
(4) $\aggbert$& 0.343& 0.962& 0.677\ /\ 0.666& 0.696&	0.805&	0.867& 0.735& 0.813& 0.860\\ \hline
 \arrayrulecolor{lightgray}
(5) $\clscondenser$& 0.335& 0.954& 0.663\ /\ 0.666& 0.701& 0.814& 0.872& 0.732& 0.812& 0.858\\
(6) $\aggcondenser$& 0.356& 0.966& 0.674\ /\ \textbf{0.697}& 0.699& 0.810& 0.873& 0.747& 0.821& 0.864 \\ 
\cline{1-10} 
(7) $\clscocondenser$& 0.352& \textbf{0.973}& 0.674\ /\ 0.684& \textbf{0.707}& \textbf{0.818}& \textbf{0.878}& 0.745& 0.819& \textbf{0.867}\\
(8) $\aggcocondenser$& \textbf{0.363}& \textbf{0.973}& 0.678\ /\ \textbf{0.697}& 0.699& 0.812& 0.875& \textbf{0.751}& \textbf{0.823}& \textbf{0.867}\\
\arrayrulecolor{black}
\hline \hline
	\end{tabular}
	}
\end{table*} 

\subsection{Datasets}

\noindent \textbf{In-Domain Evaluations.}
We evaluate in-domain retrieval effectiveness on web search and open-domain question answering.
Table~\ref{tb:data_statistics} provides statistics of the datasets.

For web search, we use the MS MARCO passage ranking dataset introduced by \citet{marco}, comprising a corpus with 8.8M passages and around 500K training queries. 
We evaluate model effectiveness on the following query sets:\ (1) MARCO Dev, 6980 queries from the development set with one relevant passage per query on average. 
Following the established procedure, we report RR@10 and R@1000 as the evaluation metrics. 
(2) TREC DL~\cite{trec19dl, trec20dl}, created by the organizers of the 2019 (2020) Deep Learning Tracks at the Text REtrieval Conferences (TRECs), where 43 (53) queries with graded relevance labels are released. 
We report nDCG@10, used by the organizers as the main metric.

For open-domain question answering, we use the Wikipedia corpus released by \citet{dpr} and conduct experiments on two query sets, Natural Questions~\citep[NQ;][]{nq} and Trivia QA~\citep[TQA;][]{tqa}. 
We directly use the training and test sets released by \citet{dpr} for training and evaluation, respectively. 
For this task, we report hit accuracy at cutoffs 5, 20, and 100, denoted R@5/20/100.

\smallskip \noindent \textbf{Zero-Shot Evaluations.}
We evaluate zero-shot retrieval effectiveness on open-domain QA with two query sets, SQuAD~\cite{squad} and Entity\-Questions~\citep[EntityQs;][]{sciavolino2021simple}, which are challenging for dense retrieval models. 
We report hit accuracy at cutoffs 20 and 100 (R@20/100). 
In addition, we use BEIR~\cite{beir}, consisting of 18 distinct IR datasets spanning diverse domains and tasks, including retrieval, question answering, fact checking, question paraphrasing, and citation prediction. 
We conduct zero-shot retrieval on 14 of the 18 datasets that are publicly available.\footnote{We exclude BioASQ, Signal-1M, TREC-NEWS, Robust04.} 
We report nDCG@10 averaged over the 14 datasets.

\subsection{Models}
\label{subsec:model}

Since our approach to text aggregation can be applied to any existing pre-trained encoder-only model, we test the effectiveness of \aggretriever on two pre-trained LM models and two further pre-trained models:\ (1) BERT~\cite{devlin2018bert};
(2) DistilBERT~\cite{distilbert}, a 6-layer transformer distilled from BERT;
(3) Condenser~\cite{condenser}, a BERT model further pre-trained with the tasks of auto-encoding and skip-connection MLM; 
and (4) coCondenser~\cite{cocondenser}, a corpus-aware Condenser combining the tasks of skip-connection MLM and an ICT variant that comes in two separate flavors, further pre-trained on the MS MARCO and Wikipedia corpora, respectively. 
All model checkpoints can be downloaded from the Hugging Face Model Hub.\footnote{\url{https://huggingface.co/models}}
We compare models fine-tuned using only the \texttt{[CLS]} vector and based on our approach with the subscripts ``CLS'' and ``AGG'', respectively, e.g., $\text{BERT}_{\text{CLS}}$ and $\text{BERT}_{\text{AGG}}$. 
In addition, we also report the effectiveness of BM25 as a reference point; these results come from the Pyserini IR toolkit~\cite{pyserini}. 

For implementation details, we refer readers to Appendix~\ref{subsec:implementation_details}.
It is worth emphasizing that in our main experiments, we do not leverage any expensive fine-tuning strategies such as hard negative mining or knowledge distillation.
Thus, we fine-tune all the DPR models under the same settings for a fair comparison.
Additional detailed comparisons are provided in Appendix~\ref{subsec:comparison_to_sota}.

\begin{table}[t!]
	\caption{In-domain retrieval effectiveness while fine-tuning models using limited training data.}
	\label{tb:train_size_result}
	\vspace{-0.2cm}
	\centering
	 \resizebox{0.5\textwidth}{!}{  
	 \setlength\tabcolsep{0.2cm}
    \begin{tabular}{l|cc|cc}
	\hline \hline
	\multicolumn{1}{c}{}&\multicolumn{4}{c}{\textbf{MARCO Dev}}\\
	\cmidrule(rl){2-5}
  \multicolumn{1}{l}{Model}& \multicolumn{2}{c}{RR@10} & \multicolumn{2}{c}{R@1K}\\ \hline
 (a) BM25& \multicolumn{2}{c|}{0.188}& \multicolumn{2}{c}{0.858} \\ \hline
\multicolumn{1}{l|}{\textbf{Train Size}}& \multicolumn{1}{c}{1K}& \multicolumn{1}{c|}{10K}& \multicolumn{1}{c}{1K}& \multicolumn{1}{c}{10K}\\ \hline
 \arrayrulecolor{lightgray}
 (1) $\clsdistilbert$& 0.145& 0.222& 0.754& 0.865\\
 (2) $\aggdistilbert$& 0.207& 0.260& 0.868& 0.905\\ \hline
 (3) $\clsbert$& 0.153& 0.230& 0.778& 0.866\\
  \arrayrulecolor{black}
 (4) $\aggbert$& 0.207& 0.258& 0.871& 0.906\\ \hline
 \arrayrulecolor{lightgray}
 (5) $\clscondenser$& 0.191& 0.259& 0.841& 0.903\\
 (6) $\aggcondenser$& 0.211& 0.258& 0.873& 0.899\\ \hline
 (7) $\clscocondenser$& \textbf{0.234}& \textbf{0.287}& \textbf{0.935}& \textbf{0.948}\\
 (8) $\aggcocondenser$& 0.209& 0.280& 0.880& 0.914\\
\arrayrulecolor{black}
	\hline \hline
	\end{tabular}
	}
\end{table} 

\section{Results}

\subsection{In-Domain Evaluations}
\begin{table*}[h]
	\caption{Near-domain zero-shot retrieval effectiveness comparisons using NQ or TQA for fine-tuning. Bold denotes the best model for that metric.}
	\label{tb:near_domain_result}
	\vspace{-0.2cm}
	\centering
	 \resizebox{0.9\textwidth}{!}{  
    \begin{tabular}{l|cc|cc|cc|cc}
    	\hline \hline
  \multicolumn{1}{l}{\textbf{Target (Source)}}& \multicolumn{2}{c}{\textbf{SQuAD (NQ)}}& \multicolumn{2}{c}{\textbf{EntityQs (NQ)}}& \multicolumn{2}{c}{\textbf{SQuAD (TQA)}}& \multicolumn{2}{c}{\textbf{EntityQs (TQA)}}\\
 \cmidrule(rl){2-3} \cmidrule(rl){4-5} \cmidrule(rl){6-7} \cmidrule(rl){8-9}
 \multicolumn{1}{l}{Model}& \multicolumn{1}{c}{R@20}& \multicolumn{1}{c}{R@100}& \multicolumn{1}{c}{R@20}& \multicolumn{1}{c}{R@100}&\multicolumn{1}{c}{R@20}& \multicolumn{1}{c}{R@100}&\multicolumn{1}{c}{R@20}& \multicolumn{1}{c}{R@100}\\ \hline
 (a) BM25& 0.712& 0.820& 0.714& 0.800& 0.712& 0.820&	0.714& 0.800\\ \hline
 \arrayrulecolor{lightgray}
(1) $\clsdistilbert$& 0.514& 0.670& 0.518& 0.650& 0.573& 0.725& 0.640& 0.751\\
 (2) $\aggdistilbert$& 0.529& 0.688& 0.564& 0.683& 0.648& 0.775& 0.713& 0.797\\ \hline
 (3) $\clsbert$& 0.512& 0.671& 0.534& 0.664& 0.581& 0.722& 0.637& 0.747\\
 (4) $\aggbert$& 0.539& 0.692& 0.562& 0.681& \textbf{0.651}&	\textbf{0.779}& 0.716& 0.798\\ 
 \arrayrulecolor{black} \hline  \arrayrulecolor{lightgray}
 (5) $\clscondenser$& 0.559& 0.705& 0.567& 0.692& 0.605& 0.742& 0.671& 0.775\\
 (6) $\aggcondenser$& 0.541& 0.692& 0.564& 0.684& 0.643& 0.772& 0.716& 0.800\\ \hline
 (7) $\clscocondenser$& \textbf{0.567}& \textbf{0.715}& 0.556& 0.684& 0.629& 0.762& 0.695& 0.791\\
 (8) $\aggcocondenser$& 0.535& 0.696& \textbf{0.584}& \textbf{0.701}& 0.646& 0.777& \textbf{0.724}& \textbf{0.804}\\ 
 \arrayrulecolor{black}
		\hline \hline
	\end{tabular}
	}
\end{table*}

\noindent \textbf{Fine-Tuning with Full Training Data.}
Table~\ref{tb:in_domain_result} compares in-domain retrieval effectiveness across the various models. 
We observe that our approach consistently improves on DistilBERT and BERT across all datasets, especially for metrics that emphasize top rankings. 
For example, $\aggdistilbert$ sees a three-point and five-point improvement over $\clsdistilbert$ on RR@10 and nDCG@10 for MS MARCO Dev and TREC DL, respectively, and over two points on R@5 for both NQ and TQA (row 2 vs 1). 
Similar trends can be observed on BERT (row 4 vs 3). 

For the further pre-trained models, we observe that both $\aggcondenser$ and $\aggcocondenser$ yield effectiveness gains on MS MARCO and TQA (rows 6 and 8), which suggests that our approach is orthogonal and additive to further pre-training methods. 
We observe that in some cases, \aggretriever using pre-trained BERT as the backbone can obtain better retrieval effectiveness than further pre-trained models that are fine-tuned only on the \texttt{[CLS]} vector. 
For example, $\aggbert$ outperforms $\clscondenser$ for MS MARCO and TQA (row 4 vs 5). 
This indicates that existing language models pre-trained on MLM can serve as an effective single-vector dense retriever, without further pre-training, using our proposed methods.
Without corpus-aware further pre-training, $\aggcondenser$ is competitive with $\clscocondenser$ on MS MARCO and TQA (row 6 vs 7).

\smallskip \noindent \textbf{Fine-Tuning with Limited Data.}
Table~\ref{tb:train_size_result} reports retrieval effectiveness when the models are fine-tuned on subsets of the MS MARCO training data. 
Specifically, we randomly sample 1K and 10K queries from the training queries and fine-tune the models on each set for 40 epochs. 
We first observe that with only 1K training queries, both $\clsdistilbert$ and $\clsbert$ underperform BM25 (rows 1, 3 vs a), while both $\aggdistilbert$ and $\aggbert$ surpass BM25 (rows 2, 4 vs a) and are on par with $\clscondenser$ (row 5), indicating that our approach successfully aggregates text information into a single vector without any further pre-training. 
We observe similar trends when fine-tuning models with 10K training queries. 

Finally, we find that $\clscocondenser$ performs the best when fine-tuning with limited training data. 
This is probably because coCondenser's further pre-training is designed for the \texttt{[CLS]} vector to learn corpus-aware signals from pseudo relevance in addition to skip-connection MLM.
Thus, the \texttt{[CLS]} vector is more ``ready'' for retrieval with small training data.

\subsection{Zero-Shot Evaluations}
\begin{table}[t!]
	\caption{Multi-domain zero-shot retrieval effectiveness comparisons using various sources for fine-tuning. Bold denotes the best model for that metric.}
	\label{tb:out_of_domain_result}
	\vspace{-0.2cm}
	\centering
	 \resizebox{0.5\textwidth}{!}{  

	 \setlength\tabcolsep{0.15cm}
    \begin{tabular}{l|c|c|c}
	\hline \hline
\multicolumn{1}{l}{Model}& \multicolumn{3}{c}{$\textbf{BEIR}$ (nDCG@10)}\\ \hline
 (a) BM25\tnote{$\ast$}& \multicolumn{3}{c}{0.430} \\ \hline
  \multicolumn{1}{l|}{\textbf{Source}}& \multicolumn{1}{c}{MARCO}& \multicolumn{1}{c}{\ \ \ \ NQ \ \ \ \ }& \multicolumn{1}{c}{\ \ \ TQA \ \ \ }\\ \hline
 \arrayrulecolor{lightgray}
 (1) $\clsdistilbert$& 0.364& 0.262& 0.266\\
 (2) $\aggdistilbert$& \textbf{0.450}& 0.277& 0.386\\ \hline
 (3) $\clsbert$& 0.382& 0.283& 0.305\\
  \arrayrulecolor{black}
 (4) $\aggbert$& 0.449& \textbf{0.299}& \textbf{0.394}\\ \hline
 \arrayrulecolor{lightgray}
 (5) $\clscondenser$& 0.393& 0.286& 0.314\\
 (6) $\aggcondenser$& 0.447&  0.295& 0.385\\ \hline
 (7) $\clscocondenser$& 0.414& 0.277&  0.307\\
 (8) $\aggcocondenser$& 0.446& 0.280& 0.376\\
\arrayrulecolor{black}
	\hline \hline
	\end{tabular}
	}
 \vspace{-0.3cm}
\end{table}

\noindent \textbf{Near-Domain Retrieval Effectiveness.}
In these experiments, we examine robustness in a zero-shot retrieval setting.
We first consider transfer to ``near-domain'' (Wikipedia) datasets, reported in Table~\ref{tb:near_domain_result}. 
Specifically, we perform retrieval on test queries from SQuAD and EntityQs using models fine-tuned on NQ or TQA. 

We see that \aggretriever with any backbone yields sizable gains over its \texttt{[CLS]} counterpart, with the exception that $\aggcondenser$ (and $\aggcocondenser$)\ underperforms $\clscondenser$ (and $\clscocondenser$)\ in SQuAD using NQ as the source (e.g., row 6 vs 5).  
It is worth mentioning that using TQA as the source, \aggretriever with any backbone is competitive with BM25 while the other \texttt{[CLS]} models still lag behind BM25 on the Entity\-Qs test queries.
Finally, we observe that models fine-tuned on TQA have better zero-shot retrieval effectiveness in near-domain datasets compared to those fine-tuned on NQ, which is also observed by \citet{spider}. 

\smallskip \noindent \textbf{Multi-Domain Retrieval Effectiveness.}
In addition, we evaluate zero-shot retrieval effectiveness on the multi-domain $\text{BEIR}$ dataset, reported in Table~\ref{tb:out_of_domain_result}. 
We evaluate the models fine-tuned on three different sources:\ MS MARCO, NQ, and TQA. 
Similarly, \aggretriever shows better zero-shot retrieval effectiveness compared to its \texttt{[CLS]} counterpart with any backbone. 
For example, our model consistently and substantially outperforms the comparable baselines using MS MARCO and TQA as the source dataset for fine-tuning. 
Although models fine-tuned on NQ show the worst zero-shot retrieval capability, \aggretriever with any backbone still slightly outperforms its \texttt{[CLS]} counterpart.
It is also worth mentioning that \aggretriever with any backbone fine-tuned on MS MARCO outperforms the strong BM25 baseline.

\begin{table}[t!]
	\caption{Fine-tuning with noisy hard negatives.}
	\label{tb:hard_neg_training}
	\vspace{-0.2cm}
	\centering
	 \resizebox{0.95\columnwidth}{!}{
    \begin{tabular}{lr|cc}
   \hline \hline
 &  \multicolumn{1}{c}{}& \multicolumn{2}{c}{\textbf{MARCO Dev}}\\
 \cmidrule(rl){3-4}
\multicolumn{2}{l}{Model \ \ \ \ \ \ \ \ \ \ \ \ \ \ \ \ \ \ \ batch size}& RR@10 & \multicolumn{1}{c}{R@1K}\\ \hline
 \multicolumn{2}{l|}{\textbf{RocketQA}~\cite{rocketqa}}& \\
 BM25 Neg.& 8K& 0.333& -\\
 + Hard Neg.& 4K& 0.260& -\\
 + Denoise& 4K& 0.364& -\\ 
 + Data Aug.& 4K& 0.370& 0.979\\ \hline
 \multicolumn{2}{l|}{\textbf{TCT}~\cite{tct}}& \\
 BM25 Neg. + KD& 96& 0.344& 0.967\\
 + Hard Neg.& 96& 0.237& 0.929\\
 + KD& 96& 0.359& 0.970\\ \hline
 \multicolumn{2}{l|}{$\text{\textbf{DistilBERT}}_{\text{\textbf{AGG}}}$}& \\ 
 BM25 Neg.& 64& 0.341& 0.960\\
 + Hard Neg.& 64& 0.360& 0.967\\ 
\hline \hline
	\end{tabular}}
\end{table}

\subsection{Fine-Tuning with Noisy Hard Negatives}

In this experiment, we use $\aggdistilbert$ to examine \aggretriever's robustness to fine-tuning with noisy hard negatives.
Following TCT~\cite{tct} and RocketQA~\cite{rocketqa}, for each query in the MS MARCO training set, we retrieve the top-200 candidates using $\aggdistilbert$ and further fine-tune the model by randomly sampling the candidates as negatives for two additional epochs using the same settings as the previous fine-tuning setup.

The results are listed in Table~\ref{tb:hard_neg_training}; we directly copy the numbers of TCT and RocketQA from the original papers. 
We notice that hard negatives reduce the effectiveness of both TCT and RocketQA since there are many false negatives in the candidates, as noted by \citet{rocketqa}. 
They address this issue using expensive training strategies:\ knowledge distillation, denoising, and cross-batch negative sampling. 
On the other hand, $\aggdistilbert$ obtains competitive retrieval effectiveness without any expensive training strategies. 
This experiment demonstrates that \aggretriever is robust and able to extract useful information when fine-tuned with hard negatives.

\begin{table}[t!]
	\caption{$\text{DistilBERT}_{\text{AGG}}$ dimensionality ablation.}
	\label{tb:dimension_ablation}
	\vspace{-0.2cm}
	\centering
	 \resizebox{0.95\columnwidth}{!}{
 \setlength\tabcolsep{0.12cm}
    \begin{tabular}{lcc|ccc}
    \hline \hline
\multicolumn{1}{c}{} &\multicolumn{2}{c}{Dim.} & \multicolumn{2}{c}{\textbf{MARCO Dev}}& \multicolumn{1}{c}{\textbf{BEIR small}} \\
 \cmidrule(rl){2-3} \cmidrule(rl){4-5} \cmidrule(rl){6-6} 
 &\texttt{[CLS]} & \multicolumn{1}{c}{$\mathbf{agg}^{\star}$} & RR@10 & R@1K& \multicolumn{1}{c}{nDCG@10}\\ \hline
 \arrayrulecolor{lightgray}
 (1)& 768 &  \ \ \ \ \ \ \ \ 0& 0.308& 0.940& 0.259\\  \hline
 (2)& 640 &  \ \ \ \ 128& 0.327& 0.954& 0.307\\  \hline
 (3)& 128 &  \ \ \ \ 640& 0.341& 0.960& 0.355\\  \hline
 \arrayrulecolor{black}
 (4)& \ \ \ \ 0 &  \ \ \ \ 768& 0.307& 0.926& 0.328\\ \hline
 \arrayrulecolor{lightgray}
 (5)& 768 &  \ \ \ \ 768& 0.350& 0.966& 0.358\\  \hline
 (6)& 128 &  \ \ \ \ 128& 0.320& 0.946& 0.300\\ \hline
 (7)& \ \ \ \ 0 & 30522& 0.345& 0.956& 0.363\\ 
 \arrayrulecolor{black}
\hline \hline
	\end{tabular}}
\end{table}
 
\begin{table*}[t!]
	\caption{$\text{DistilBERT}_{\text{AGG}}$ text aggregation ablation. We project \texttt{[CLS]} to 128 dimensions and concatenate with a 640-dimensional embedding pooled and pruned using different strategies. AVERAGE denotes average pooling over all 768-dimensional contextualized token embeddings other than \texttt{[CLS]}.}
	\label{tb:pooling_stage_ablation}
	\vspace{-0.2cm}
	\centering
	 \resizebox{0.75\textwidth}{!}{
    \begin{tabular}{ccc|c|ccc}
\hline \hline
  \multicolumn{1}{c}{}& \multicolumn{2}{c|}{Pooling}& \multicolumn{1}{c|}{Pruning} & \multicolumn{2}{c}{\textbf{MARCO Dev}}& \multicolumn{1}{c}{\textbf{BEIR small}} \\
 \cmidrule(rl){2-3} \cmidrule(rl){5-6} \cmidrule(rl){7-7}  
 & MLM& \multicolumn{1}{c|}{Weight}& & RR@10& R@1K& \multicolumn{1}{c}{nDCG@10} \\
 \hline
\arrayrulecolor{lightgray}
(1)& \cmark& \cmark& full aggregation& 0.341& 0.960& 0.355\\ \hline
(2)& \cmark& \xmark& full aggregation& 0.308& 0.937& 0.308\\ \hline
\arrayrulecolor{black}
(3)& \xmark& \cmark& full aggregation& 0.332& 0.953& 0.355\\  \hline
\arrayrulecolor{lightgray}
(4)& \cmark& \cmark& semi aggregation& 0.341& 0.960& 0.322\\ \hline
(5)& \cmark& \cmark& $\texttt{linear}(|\text{V}_{\text{BERT}}| \rightarrow 640)$& 0.327& 0.959& 0.313\\
\arrayrulecolor{black}
\hline
(6)& \multicolumn{2}{c|}{AVERAGE}& $\texttt{linear}(768 \rightarrow 640)$ \ \ \ \  & 0.300& 0.933& 0.270\\
(7)& \multicolumn{2}{c|}{RepBERT~\cite{repbert}}& -& 0.306& 0.942& 0.264\\
\arrayrulecolor{black}
\hline \hline
	\end{tabular}
 }
\end{table*}

\subsection{Ablation Study}
\label{subsec:ablation}

In this experiment, we use $\aggdistilbert$ fine-tuned on the MS MARCO dataset to conduct an ablation study. 
In addition to MARCO Dev, to understand the zero-shot effectiveness of each condition, we conduct retrieval on a subset of BEIR (denoted BEIR small), consisting of five datasets from different domains:\ NFCorpus, FiQA, Argu\-Ana, SCIDOCS, and SciFact. 
We report nDCG@10 averaged over these five datasets. 

\smallskip \noindent \textbf{Dimensionality Ablation.} 
We first study the effects of dimensionality on the \texttt{[CLS]} and $\mathbf{agg}^{\star}$ vectors in Table~\ref{tb:dimension_ablation}. 
We find that \texttt{[CLS]} alone slightly outperforms $\mathbf{agg}^{\star}$ alone (row 1 vs 4) on in-domain evaluation while the reverse trend is seen on zero-shot evaluation. 
This observation indicates that the \texttt{[CLS]} and $\mathbf{agg}^{\star}$ vectors encode text in different ways and that combining them further improves retrieval effectiveness (row 5). 
Compared to \texttt{[CLS]} alone and $\mathbf{agg}^{\star}$ alone, we still see a slight improvement for in-domain evaluation at 256 dimensions (row 6 vs 1 and 4). 
Holding the number of dimensions constant (rows 1--4), the best condition (row 3) indicates that the $\mathbf{agg}^{\star}$ vector requires more space than the $\texttt{[CLS]}$ vector. 

Finally, we report the retrieval effectiveness of the original wordpiece lexical representations before pruning (row 7), which can be considered the effectiveness upper bound of $\mathbf{agg}^{\star}$. 
Although $\mathbf{agg}^{\star}$ with 768 dimensions has lower effectiveness (row 4 vs 7), combined with \texttt{[CLS]}, \aggretriever reduces the gap (rows 3, 5 vs 7), with better retrieval efficiency in terms of smaller index size and lower retrieval latency. 
For example, on the MS MARCO dataset, representing each passage as a 768-dimensional vector in a Faiss Flat index with 32 (16) bits requires 26 (13) GB and 100 ms/q retrieval latency on a single V100 GPU, while the 30522-dimensional vectors (without pruning) require around 40 times more index storage and are not practical for end-to-end retrieval.

\smallskip \noindent \textbf{Pooling Stage Ablation.} 
In the second ablation experiment, we fix $\texttt{[CLS]}$ and $\mathbf{agg}^{\star}$ to 128 and 640 dimensions, respectively, and compare different designs of the pooling stage to form $\mathbf{agg}^{\star}$, as discussed in Section~\ref{subsec:text_aggregation:pooling}. 
The results are reported in the first main block of Table~\ref{tb:pooling_stage_ablation}; row 1 is our default condition. 
In row 2, we remove the term importance component and assign a term weight of one for weighted max pooling. 
A substantial drop in retrieval effectiveness can be observed. 
In row 3, we remove MLM projection and represent each query (or passage) token with the $30522$-dimensional indicator vector in Eq.~(\ref{eq:mlm_project}); that is, $\mathbf{p}_{q_i} = x_j \in \{0,1\}^{|\text{V}_{\text{BERT}}|}$ for $j \in \{\text{token\_id}(q_i)\}$.
We notice that skipping the MLM projector modestly harms retrieval effectiveness. 
This means that most textual information can be captured without the MLM projector, but it {\it does} help.
This is sensible since the $30522$-dimensional indicator vector still retains each original query (or passage) term. 
A comparison of row 2 and row 3 shows that learned term weights for each token are more important than the term semantic distribution (projected by MLM) over the wordpiece vocabulary.

\smallskip \noindent \textbf{Pruning Stage Ablation.}
In the second main block of Table~\ref{tb:pooling_stage_ablation}, we study the effects of pruning wordpiece lexical representations on \aggretriever. 
For example, we semi-aggregate (linearly project) the lexical representations into 640-dimensional dense vectors, as shown in row 4 (5). 
We observe that our non-parametric pruning approaches are better than the learned ones (rows 1, 4 vs 5). 
Although $\mathbf{agg}^+$ shows the same retrieval effectiveness as $\mathbf{agg}^{\star}$ on in-domain evaluation, a substantial drop can be observed on out-of-domain evaluation (row 1 vs 4). 
This result demonstrates that our fully aggregated representations better preserve information from lexical representations and appear to be more robust to domain shifts.

We observe that directly projecting averaged contextualized embedding (excluding the \texttt{[CLS]}), denoted AVERAGE, into 640 dimensions, and then concatenating with \texttt{[CLS]} (row 6), does not perform well, indicating that projecting contextualized token embeddings into the high-dimensional wordpiece lexical space before pooling is key to preserving lexical information. 
Finally, we also try average pooling over all contextualized embeddings (including the \texttt{[CLS]}), which corresponds to RepBERT~\cite{repbert}.
This yields negligible effectiveness difference from AVERAGE concatenated with $\texttt{[CLS]}$ (row 7 vs 6); i.e., 0.306 (RR@10) and 0.264 (nDCG@10) on MARCO dev and BEIR small, respectively.

\begin{figure}[t!]
    \centering
    \resizebox{1\columnwidth}{!}{
        \includegraphics{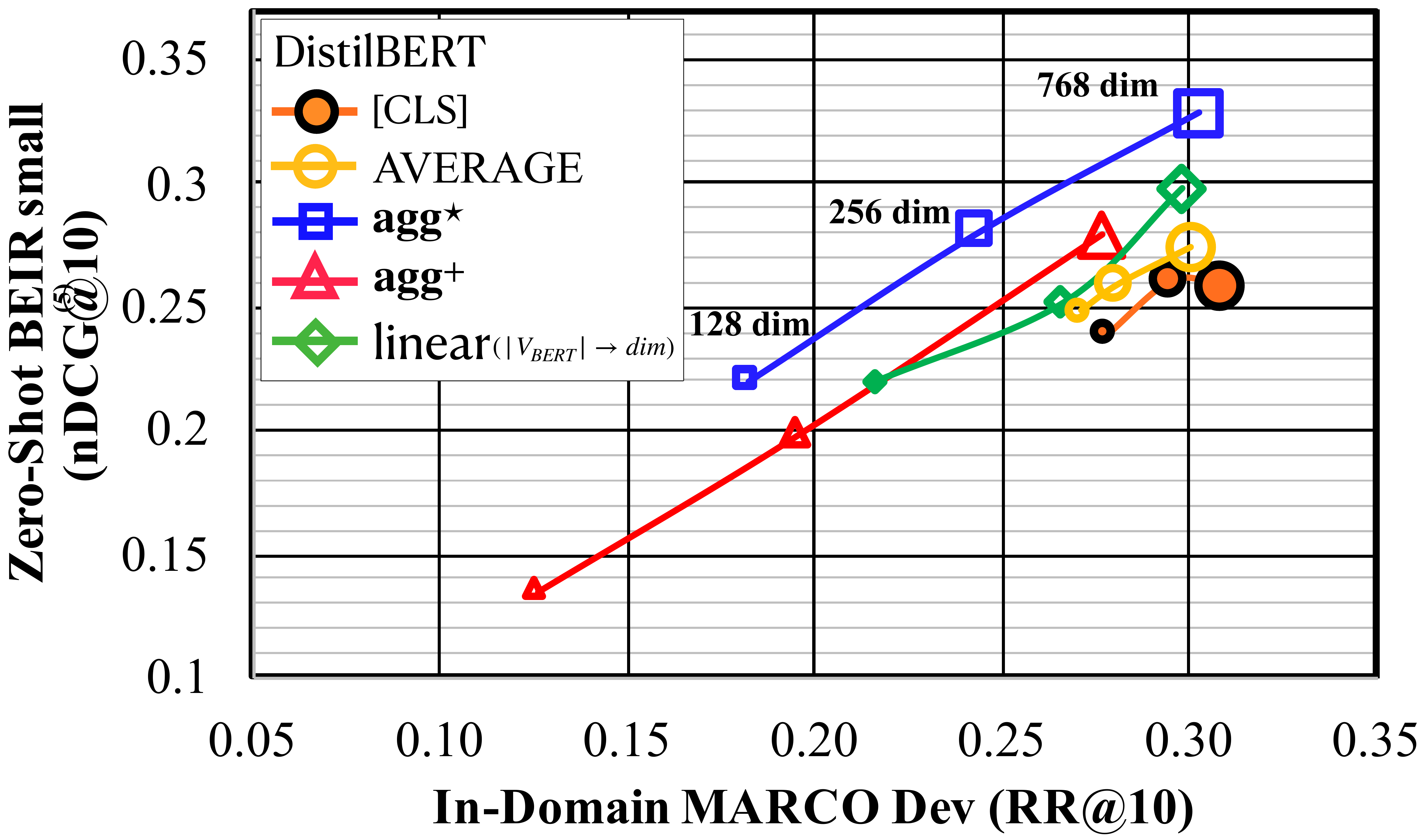}
    }
    \caption{In-domain versus zero-shot effectiveness comparisons between textual representations under different numbers of dimensions.}
    \label{fig:rep_comparison}
\end{figure}

To further understand the differences between pruned lexical representations (rows 1, 4, 5 in Table~\ref{tb:pooling_stage_ablation}), we fine-tune DistilBERT using each representation alone (without using \texttt{[CLS]}) with 128, 256, and 768 dimensions on the MS MARCO dataset and compare their retrieval effectiveness on MS MARCO Dev and BEIR small in Fig.~\ref{fig:rep_comparison}. 
We observe that $\mathbf{agg}^{\star}$ performs better than $\mathbf{agg}^+$ under all conditions, demonstrating that distributing representations to the full vector space can mitigate the problem of term misalignment (rectangles vs triangles) mentioned in Section~\ref{subsec:text_aggregation:pruning}, especially when the number of dimensions is small. 
Although the linearly projected lexical representations (diamonds) show better in-domain retrieval effectiveness than our non-parametric pruning approaches ($\mathbf{agg}^{+}$ and $\mathbf{agg}^{\star}$) with 128 and 256 dimensions, $\mathbf{agg}^{\star}$ still exhibits better zero-shot retrieval effectiveness. 
This indicates that the learned linear projector helps compress textual information into low-dimensional space in a way that is biased toward the training data.  

In addition, in Fig.~\ref{fig:rep_comparison}, we also show the retrieval effectiveness of \texttt{[CLS]} and AVERAGE (solid and hollow circles) as comparisons.
We observe that although all 768-dimensional textual representations reach similar in-domain retrieval effectiveness, \texttt{[CLS]} and AVERAGE show poor zero-shot retrieval effectiveness on BEIR small compared to the other models pruned from 30K-dimensional lexical representations. 
We hypothesize that \texttt{[CLS]} and AVERAGE capture textual information in a different manner than our lexical representations. 
This explains why fusing \texttt{[CLS]} with pruned lexical representations performs better than AVERAGE (rows 1, 4, 5 vs 6 in Table~\ref{tb:pooling_stage_ablation}).

However, \texttt{[CLS]} and AVERAGE do not exhibit much retrieval effectiveness drop on both in-domain and zero-shot evaluations when reducing the number of dimensions.
This is probably because lexical representations contain fine-grained textual information in 30K-dimensional lexical space while \texttt{[CLS]} and AVERAGE embeddings capture high-level textual information in low-dimensional semantic space. 
This result also explains the optimal balance in Table~\ref{tb:dimension_ablation}, where $\mathbf{agg}^{\star}$ requires more space than \texttt{[CLS]} when restricting the total vector dimension to 768.

\begin{table}[t!]
	\caption{Query encoding latency comparisons.}
	\label{tb:query_encoding_latency}
	\vspace{-0.2cm}
	\centering
	 \resizebox{0.49\textwidth}{!}{
	\setlength\tabcolsep{0.1cm}
    \begin{tabular}{cr|c|c}
   \hline \hline
    & \multicolumn{1}{c}{}& \multicolumn{2}{c}{\textbf{latency} ($1^{\text{st}}\ /\ 50^{\text{th}}\ /\ 99^{\text{th}}$ perc.)}\\ 
    \cmidrule(rl){3-4}
  & \multicolumn{1}{c}{}& \multicolumn{1}{c}{CPU}& GPU\\
  \hline
  (1)& $\clsdistilbert$ \ & \ \ 93\ /\ 103\ /\ 122 ms& 15\ /\ 16\ /\ 18 ms\\
  \arrayrulecolor{lightgray}
  \hline
  (2)& $\aggdistilbert$& 155\ /\ 163\ /\ 191 ms& 18\ /\ 19\ /\ 24 ms\\ 
  \arrayrulecolor{black}
  (3)& w/o MLM& 103\ /\ 109\ /\ 138 ms& 16\ /\ 19\ /\ 20 ms\\
  \arrayrulecolor{black}
\hline \hline
	\end{tabular}}
\end{table}

\subsection{Query Encoding Latency}
\label{subsec:query_encoding_latency}

Although different single-vector dense retrievers with the same vector dimensionality have similar retrieval latency under the same software and environment when performing top-$k$ retrieval, query encoding latency is also an important component to consider.
In this experiment, we compare the query encoding latency of $\aggdistilbert$ and $\clsdistilbert$. 
We measure the time required to encode the 6980 queries from MARCO Dev with batch size one on the CPU and GPU, using one thread on a Linux machine with a 2.2 GHz Intel Xeon Silver 4210 CPU and a single Tesla V100 GPU (32GB), respectively.
We report the latency at $1^{\text{th}}$, $50^{\text{th}}$ and $99^{\text{th}}$ percentiles in Table~\ref{tb:query_encoding_latency}. 

We observe that query encoding with \aggretriever is slightly slower than its \texttt{[CLS]} counterpart on the GPU (row 2 vs 1). 
On the CPU, the gap is much larger, especially for tail queries. 
However, from row 3 (the same condition as row 3 in Table~\ref{tb:pooling_stage_ablation}), we see that skipping the MLM head projection step reduces the query encoding latency with only a small retrieval effectiveness loss. 
For a real-world application, this might be a sensible option, bringing query encoding latency roughly in line with the \texttt{[CLS]}-only model.

\subsection{Comparison with Sparse Retrievers}
\begin{table}[t!]
	\caption{Comparison with sparse retrievers.}
	\label{tb:comparison_sparse}
	\vspace{-0.2cm}
	\centering
	 \resizebox{1\columnwidth}{!}{
	 \begin{threeparttable}
 \setlength\tabcolsep{0.18cm}
    \begin{tabular}{cl|ccc}
    \hline \hline
& & \multicolumn{2}{c}{\textbf{MARCO Dev}}& \multicolumn{1}{c}{\textbf{$\text{BEIR}$}}  \\
 \cmidrule(rl){3-4} \cmidrule(rl){5-5}
& & RR@10 & R@1K& \multicolumn{1}{c}{nDCG@10}\\ \hline
 \arrayrulecolor{lightgray}
(1)& $\clsdistilbert$& 0.308& 0.940& 0.364\\ \hline 
  \arrayrulecolor{black}
(2)& $\aggdistilbert$& 0.341& 0.960& 0.450\\  
(3)& w/o MLM& 0.332& 0.953& 0.445\\  \hline
(4)& SPLADE-max& 0.340& 0.965& 0.447\\
(5)& w/o MLM\tnote{$\ast$}& 0.315& 0.924& 0.441\\
 \arrayrulecolor{black}
\hline \hline
	\end{tabular}
		\begin{tablenotes}
	\item[$\ast$] uniCOIL w/o expansion~\cite{unicoil} can be considered a variant of SPLADE-max w/o MLM.
    \end{tablenotes}
    \end{threeparttable}
	}
  \vspace{-0.3cm}
\end{table}

In our final set of experiments, we compare \aggretriever and sparse retrievers since we borrow ideas from existing learned sparse retrieval models such as SPLADE-max~\cite{splade-v2}, which uses a different activation function after the MLM projector and adds sparsity regularization to generate sparse lexical representations for inverted indexes.
For comparison to a sparse retriever without MLM projection, we use uniCOIL without expansions from T5~\cite{doctttttquery}.
Both models are fine-tuned on MS MARCO with BM25 negatives; thus, they represent reasonably fair comparisons to $\aggdistilbert$ and its variant without MLM, respectively (although uniCOIL uses BERT as a backbone).
We index and evaluate SPLADE-max and uniCOIL using the code provided by \citet{splade-v2}\footnote{\url{https://github.com/naver/splade}} and Pyserini~\cite{pyserini}, respectively.\footnote{Note that the BEIR figures for SPLADE-max reported in \citet{splade-v2} do not include CQADupStack and use Tóuche-2020 {\small(v1}) instead of Tóuche-2020 {\small(v2)}.}

Results are shown in 	Table~\ref{tb:comparison_sparse}.
We first observe that $\clsdistilbert$ shows competitive in-domain retrieval effectiveness but underperforms sparse retrievers on out-of-domain evaluations (row 1 vs 5). 
This indicates that sparse retrieval using lexical matching has better generalization across retrieval tasks than dense retrieval with \texttt{[CLS]} alone. 
On the other hand, $\aggdistilbert$ and its variant show equally good generalization capability compared to the sparse retrievers (rows 2, 3 vs 4, 5). 
We attribute the transferability of \aggretriever to $\mathbf{agg}^{\star}$, which effectively aggregates and preserves information from wordpiece lexical representations.

Finally, we observe that without the MLM projector, the effectiveness of the sparse retrievers degrades, especially on in-domain evaluation (row 4 vs 5), while $\mathbf{agg}^{\star}$ only sees a slight degradation (row 2 vs 3). 
We hypothesize that the MLM projector helps sparse retrievers learn semantic matching as well as exact term matching.
In contrast, \aggretriever can still learn semantic matching, even without the MLM projector, because it benefits from fusion with the \texttt{[CLS]} vector. 

\section{Related Work}

\noindent \textbf{Dense Retrieval.}
The most related line of research to our own work is the literature on how to effectively fine-tune a single-vector dense retriever. 
On the one hand, some researchers propose computationally expensive fine-tuning techniques such as hard negative mining strategies~\cite{xiong2020approximate, star}, knowledge distillation~\cite{tct, tasb}, or their combination~\cite{rocketqa}. 
On the other hand, others leverage further pre-training to improve the subsequent fine-tuning~\cite{ict, simcse, seed-encoder, condenser, contriever, cocondenser, retromae}. 
As far as we are aware, our work is the first to discuss how to fine-tune dense retrieval models to effectively aggregate textual information from the pre-trained MLM head rather than directly using the \texttt{[CLS]} vector or contextualized embeddings from max or average pooling~\cite{sentence-bert}.

\smallskip \noindent \textbf{Sparse Retrieval.}
Previous work \cite{sparterm, deepimpact, splade, unicoil} has demonstrated that projecting contextualized token embeddings into a high-dimensional vector in the wordpiece vocabulary space is an effective way to represent token-level information from transformers for lexical matching. 
These models directly feed the high-dimensional vectors into an inverted index for retrieval.
Thus, sparsity control for effectiveness--efficiency tradeoffs involves additional considerations~\cite{Mackenzie_etal_arXiv2021}. 
In contrast, our approach converts high-dimensional vectors into low-dimensional ones where top-$k$ retrieval can be performed directly using ANN search libraries~\cite{scann, faiss}.

\smallskip \noindent \textbf{Hybrid Retrieval.}
Our work can be characterized as hybrid since we ``fuse'' semantic and lexical representations into a single dense vector.
Recent work~\cite{coil, colberter, unifier, dhr} proposes to jointly train \texttt{[CLS]} and token-level representations for semantic and lexical matching, respectively. 
The two kinds of representations require different implementations for top-$k$ retrieval, so multiple software stacks are required to perform retrieval.
In contrast, our representations retain the best of semantic and lexical matching, but entirely as dense vectors.
Thus, retrieval can be performed in a simple execution environment.

\section{Conclusion and Future Work}

In this paper, 
we present \aggretriever, a single-vector dense retrieval model that exploits all contextualized token embeddings from the input to BERT. 
We introduce a simple approach to aggregate the contextualized token embeddings into a dense vector, $\mathbf{agg}^{\star}$. 
Experiments show that $\mathbf{agg}^{\star}$ combined with the standard \texttt{[CLS]} vector achieves better retrieval effectiveness than using the \texttt{[CLS]} vector alone for both in-domain and zero-shot evaluations. 
Our work demonstrates that MLM pre-trained transformers can be fine-tuned into effective dense retrievers without further pre-training or expensive fine-tuning strategies.

Our work leads to a few open questions for future research:\ (1) Since we have demonstrated that \aggretriever still benefits from further pre-training, can we design additional pre-training tasks tailored directly to our model?
The design of these tasks, of course, needs to be mindful of the computational costs.
(2) 
Can we apply current state-of-the-art compression techniques to \aggretriever? 
\citet{jpq, RepCONC} has shown that 768-dimensional dense representations can be effectively compressed into much smaller vectors.
However, it is still unknown if these techniques can be applied to \aggretriever to retain both in-domain and zero-shot retrieval effectiveness. 
(3) 
Finally, can we apply \aggretriever to multi-lingual retrieval?
Since in a multi-lingual BERT model, the MLM head can project into tokens in multiple languages, we can envision a natural extension. 
However, as shown in Section~\ref{subsec:query_encoding_latency}, MLM projection is expensive, and the issue becomes worse when using a pre-trained multi-lingual model since the vocabulary size is usually even larger.

\section*{Acknowledgements} 

This research was supported in part by the Canada First Research Excellence Fund and the Natural Sciences and Engineering Research Council (NSERC) of Canada. 
We thank the anonymous referees who provided useful feedback to improve this work.

\bibliographystyle{acl_natbib}
\bibliography{paper.bib}

\clearpage

\appendix
\appendix
\section{Appendix}

\subsection{Implementation Details}
\label{subsec:implementation_details}

We implement our models using Tevatron~\cite{tevatron} and apply its default training settings in most tasks. 
For MS MARCO, we train models for three epochs with a learning rate $5e-6$, and for each batch, we include 8 queries. 
Each of the queries is paired with a randomly sampled positive passage and 7 negative passages mined using BM25. 
The maximum query and passage lengths are set to 32 and 128, respectively.
Note that we use the official training set and corpus\footnote{\url{https://microsoft.github.io/msmarco/TREC-Deep-Learning-2019}} instead of the ones in Tevatron, which are further processed by \citet{rocketqa}. 
For open-domain QA, we follow the original settings used by \citet{dpr} except for two modifications:\ (1) we use shared instead of independent weights between the query and passage encoders; (2) we set the maximum query and passage lengths to 32 and 156 for faster fine-tuning and inference. 
Note that we use one and four Tesla V100 GPUs (32GB) for model fine-tuning on MS MARCO and open-domain QA, respectively.
For BEIR evaluation, we use the APIs provided by \citet{beir} and set maximum query and passage input lengths to 512.\footnote{\url{https://github.com/beir-cellar/beir}}

\subsection{Comparison with Existing DPR Models}
\label{subsec:comparison_to_sota}

\begin{table}[t]
	\caption{Comparisons with existing DPR models.}
	\label{tb:advanced_comparison}
	\centering
	 \resizebox{0.48\textwidth}{!}{  
	 \begin{threeparttable}
	  	 \setlength\tabcolsep{3pt}
    \begin{tabular}{lccc!{\color{lightgray}\vrule}cccc}
	\hline \hline
&\multicolumn{3}{c!{\color{lightgray}\vrule}}{w/o pre-training}& \multicolumn{4}{c}{w/ pre-training}\\
\hline
&\rotatebox[origin=c]{295}{$\aggdistilbert$}&
 \rotatebox[origin=c]{295}{TAS-B}&
\rotatebox[origin=c]{295}{CL-DRD}&
\rotatebox[origin=c]{295}{$\aggcocondenser$}&
\rotatebox[origin=c]{295}{coCondenser}&
 \rotatebox[origin=c]{295}{Contriever}&
 \rotatebox[origin=c]{295}{GTR-Base}\\
\hline
  \arrayrulecolor{lightgray}
 \multicolumn{1}{l}{KD}&  \xmark  & \cmark & \cmark &  \xmark &  \xmark &  \xmark& \cmark\\
\hline
 \multicolumn{1}{l}{HNM}& \xmark  & \xmark & \cmark& \xmark & \cmark & \cmark & \cmark\\
 \hline
 \multicolumn{1}{l}{batch size >1K} & \xmark  & \xmark & \xmark& \xmark & \xmark & \cmark & \cmark\\
 \arrayrulecolor{black}
\hline
MARCO&\multicolumn{7}{c}{RR@10} \\
 \arrayrulecolor{lightgray}
\hline
\textbf{Dev} & 0.341& 0.344& 0.381& 0.363& 0.382$\tnote{$\ast$}$& 0.341& 0.366\\
\arrayrulecolor{black}
\hline
\multicolumn{1}{l}{BEIR}& \multicolumn{7}{c}{nDCG@10} \\
\arrayrulecolor{lightgray}
\hline
\textbf{TREC-COVID}& 0.661& 0.481& 0.584& 0.751& 0.712& 0.596& 0.539\\
\textbf{NFCorpus}& 0.297& 0.319& 0.315& 0.323& 0.325& 0.328& 0.308\\
\textbf{NQ}& 0.474& 0.463& 0.500& 0.490& 0.487& 0.498& 0.495\\
\textbf{HotpotQA}& 0.616& 0.584& 0.589& 0.609& 0.563& 0.638& 0.535\\
\textbf{FiQA-2018}& 0.292& 0.300& 0.308& 0.305& 0.276& 0.329& 0.349\\
\textbf{ArguAna}& 0.417& 0.429& 0.413& 0.438& 0.299& 0.446& 0.511\\
\textbf{Tóuche-2020 {\small(v2)}}& 0.263& 0.162& 0.203& 0.213& 0.191& 0.230& 0.205 \\
\textbf{Quora}& 0.834& 0.835& 0.826& 0.851& 0.856&	0.865& 0.881\\
\textbf{DBPedia}& 0.362& 0.384& 0.381& 0.380& 0.363& 0.413& 0.347\\
\textbf{SCIDOCS}& 0.138& 0.149& 0.146& 0.143& 0.137& 0.165& 0.149 \\
\textbf{FEVER}& 0.781& 0.700& 0.734& 0.600& 0.495& 0.758& 0.660\\
\textbf{Climate-FEVER}& 0.210& 0.228& 0.204& 0.155& 0.144& 0.237& 0.241\\
\textbf{SciFact}& 0.630& 0.643& 0.621& 0.650& 0.615& 0.677& 0.600\\
\textbf{CQADupStack}& 0.318& 0.314& 0.325& 0.338& 0.320& 0.345& 0.357\\
\arrayrulecolor{lightgray}
\hline
Avg.nDCG@10& 0.450& 0.428& 0.439& 0.446& 0.413& 0.466& 0.441\\
\arrayrulecolor{black}
	\hline \hline
	\end{tabular}
		\begin{tablenotes}
	\item[$\ast$] These numbers are not comparable due to the use of a ``non-standard'' MS MARCO passage corpus that has been augmented with title. 
    \end{tablenotes}
    \end{threeparttable}
	}
 \vspace{-0.3cm}
\end{table}

Table~\ref{tb:advanced_comparison} compares \aggretriever with existing dense retrievers fine-tuned with more expensive strategies; i.e., cross-encoder knowledge distillation (KD), hard negative mining (HNM), and large in-batch negatives, 
on both in-domain and out-of-domain evaluations. 
The two baseline models without further pre-training are:\ (1) TAS-B~\cite{tasb}, which distills ColBERT and a cross-encoder to DPR with an efficient topic-aware sampling strategy; (2) CL-DRD~\cite{cldrd}, which further improves TAS-B by combining curriculum learning, HNM, and cross-encoder KD. 
Three models with further pre-training are included:\ (1) coCondenser~\cite{cocondenser}, already discussed in Section~\ref{subsec:model}; (2) Contriever~\cite{contriever}, which leverages pre-training by combining advanced contrastive learning techniques with an Inverse Cloze Task (ICT) variant; (3) GTR-Base~\cite{gtr}, which trains a T5-Base encoder model that combines pre-training, KD, and HNM. 
For TAS-B, Contriever, and GTR-Base, we directly copy numbers from \citet{contriever} and \citet{gtr}, respectively. 
For CL-DRD\footnote{\url{https://github.com/HansiZeng/CL-DRD}} and coCondenser,\footnote{\url{https://huggingface.co/Luyu/co-condenser-marco-retriever}} we use the models provided by the authors to conduct in-domain and out-of-domain evaluations ourselves. 
Note that the coCondenser model provided by the authors is fine-tuned in another round with self-mined hard negatives.
Furthermore, they use a ``non-standard'' MS MARCO corpus where each passage is concatenated with a title; thus, the MS MARCO Dev results are different from the values for $\clscocondenser$ reported in Table~\ref{tb:in_domain_result}. 

First, we observe that $\aggdistilbert$ is not only competitive with TAS-B on in-domain evaluation but also outperforms both TAS-B and CL-DRD on out-of-domain evaluation, without needing supervision from an expensive cross-encoder teacher. 
Secondly, Contriever yields the best out-of-domain results at the cost of in-domain effectiveness.
On the other hand, $\aggcocondenser$ reaches the same level of retrieval effectiveness as GTR-Base without leveraging any expensive fine-tuning strategies. 
Fine-tuning \aggretriever with KD, HNM, and large batch size is possible to further improve retrieval effectiveness, but these techniques are orthogonal to our proposed model.

\end{document}